\documentclass[%
 reprint,
 amsmath,amssymb,
 aps,
]{revtex4-2}

\usepackage{graphicx}
\usepackage{dcolumn}
\usepackage{bm}
\usepackage{hyperref}

\begin{document}

\preprint{APS/123-QED}

\title{Hybrid ab initio and empirical machine learning models for the potential energy surface}

\author{Pablo Peña-Cano}
 \affiliation{CIC nanoGUNE, Tolosa Hiribidea 76, Donostia 20018, San Sebastian, Spain}
\author{Pablo M. Piaggi}%
 \email{pm.piaggi@nanogune.eu}
 \affiliation{CIC nanoGUNE, Tolosa Hiribidea 76, Donostia 20018, San Sebastian, Spain}
\affiliation{Ikerbasque, Basque Foundation for Science, Bilbao 48013, Spain}%

\date{\today}

\begin{abstract}
We propose a methodology to generate hybrid machine learning models for the potential energy surface trained simultaneously on data from ab initio electronic structure calculations and on thermodynamic and/or structural observables from experiment. The approach is based on the use of a loss function that includes the mean square error of observables with respect to their experimental values, in addition to the usual terms involving the mean square error of the energies and forces with respect to ab initio data. We employ a reweighting procedure that allows for the calculation of ensemble averages of observables during training for arbitrary values of the model parameters and on the fly. The method is general and can be applied to any set of static observables. We illustrate the usefulness of this approach by applying it to the generation of hybrid models for liquid water that reproduce accurately the experimental density maximum, the density isobar at 1 bar, and the radial distribution function in molecular dynamics simulations.
\end{abstract}

\maketitle


Molecular dynamics (MD) simulations have become an essential tool to complement experiment and provide insight into a myriad of different processes in chemistry and physics, from protein folding to crystallization~\cite{frenkel-2023}. Models for the potential energy surface (PES), also called interatomic potentials, are at the heart of this technique and the properties predicted from the simulation depend crucially on them. Until recently, models for the PES were frequently based on simple and physically-inspired functional forms with a small number of adjustable parameters fit to selected experimental properties. These empirical models were applied with great success to many systems, yet they also have a number of disadvantages, such as limited transferability to different thermodynamic conditions, absence of reactivity and polarization, and insufficient expressiveness to yield good agreement across a large number of experimental properties~\cite{abascal-2005,blazquez2023scaled,wang-2004}.

In the last five years, machine learning (ML) models for the PES based purely on data from \textit{ab initio} electronic structure calculations emerged as a highly-promising alternative to their empirical counterpart~\cite{behler-2007,zhang-2018,deringer2021gaussian}. ML models for the PES are trained on a large database of energies and forces derived usually from Kohn-Sham Density-Functional Theory (DFT)~\cite{kohn-1965}, although more sophisticated electronic-structure methods are becoming increasingly common~\cite{lan2021simulating,daru2022coupled}. Such models are nowadays widely used and highly appreciated for their predictive power across thermodynamic conditions as well as for their ability to capture bond forming and breaking (chemical reactivity), complex many body interactions, and polarization effects, which are difficult to include in empirical models for the PES. However, when properties such as densities or melting temperatures are considered, ML models for the PES often have large errors with respect to experiment, which can reach up to 10 \%~\cite{piaggi2021phase,de-hijes-2024}. These errors stem mostly from the approximations made in the electronic structure methods, although lack of inclusion of nuclear quantum effects is another source of error in systems with light species such as hydrogen or lithium~\cite{ceriotti2016nuclear}. Even though increasingly accurate and affordable electronic structure methods are being developed~\cite{ye2024periodic}, the pace of advance is relatively slow and most current ML models suffer from the limitations described above. Thus, there is urgent need for ML models for the PES that retain the benefits of \textit{ab initio} calculations, but that also provide excellent agreement with selected experimental observables.

In this Letter, we present a methodology to generate hybrid ML models for the PES trained both on information from \textit{ab initio} calculations and static (i.e., thermodynamic or structural) observables from experiment. We show the usefulness of our approach by applying it to the challenging case of liquid water, where semi-local DFT often has errors in the 5-10 \% range for many observables~\cite{piaggi2021phase,de-hijes-2024}. 

%
We consider a model for the PES $E\left(\mathbf{R};\boldsymbol{\theta}\right)$, where $\mathbf{R}$ are the atomic coordinates and $\boldsymbol{\theta}$ are learnable parameters. The usual procedure for training such a model from \textit{ab initio} calculations is based on the minimization of a loss function of the form,
\begin{align}
  \mathcal{L}_0(\boldsymbol{\theta}) =
  & p_E \sum_{i \in \mathcal{B}}\left(E\left(\mathbf{R}^i;\boldsymbol{\theta}\right) - E_\mathrm{ref}^i\right)^2 + \nonumber \\ 
  & p_f \sum_{i \in \mathcal{B}} \left\lVert \mathbf{f}\left(\mathbf{R}^i;\boldsymbol{\theta} \right) - \mathbf{f}_\mathrm{ref}^i \right\rVert^2,
  \label{eq:loss-ef}
\end{align}
where $\mathbf{R}^i$ are the atomic coordinates of the $i$-th configuration in a mini-batch $\mathcal{B}$ (subset of the training set), $E\left(\mathbf{R}^i;\boldsymbol{\theta}\right)$ and $\mathbf{f}\left(\mathbf{R}^i;\boldsymbol{\theta} \right) = - \nabla E\left(\mathbf{R}^i;\boldsymbol{\theta}\right)$ are the energies and forces predicted for $\mathbf{R}^i$ by the model, $E_\mathrm{ref}^i$ and  $\mathbf{f}_\mathrm{ref}^i$ are the \textit{ab initio} reference energies and forces for $\mathbf{R}^i$, and $p_E$ and $p_f$ are tunable prefactors that control the relative importance of each term.

We now focus on how to improve the agreement of the ensemble average of a specific observable $O(\mathbf{R})$ with respect to its experimental value.
We restrict our analysis to static observables, i.e., those that depend only on the atomic coordinates $\mathbf{R}$.
Such an observable may be a scalar (e.g., density), a vector (e.g., radial distribution function, which is not formally a vector, but it can be conveniently represented numerically as such), or even a tensor (e.g., polarizability).
The average of this observable in the isothermal-isobaric ensemble is,
\begin{equation}
  \langle O \rangle (\boldsymbol{\theta}) = \dfrac{1}{Z}\int d\mathbf{R} \: O(\mathbf{R}) e^{-\beta (E(\mathbf{R}; \boldsymbol{\theta}) + PV)},
\end{equation}
where $P$ the pressure, $V$ the volume, $\beta$ the inverse temperature and $Z$ the appropriate partition function.
We have chosen the isothermal-isobaric ensemble because it represents the most common experimental condition, but other choices are also possible.
The ensemble average can be computed by running an MD simulation, evaluating the observable at different time steps, and using,
\begin{equation}
  \langle O \rangle (\boldsymbol{\theta}) = \dfrac{1}{N} \sum_{i=1}^{N} O_i,
  \label{eq:ensemble-obs}
\end{equation}
where $O_i$ is the $i$-th observation and $N$ is the total number of observations.

To achieve our goal, we propose to modify the loss function in Eq.~\eqref{eq:loss-ef} by adding a new term that drives the observable toward its experimental value,
\begin{equation}
  \mathcal{L}(\boldsymbol{\theta}) = \mathcal{L}_0(\boldsymbol{\theta}) +
  p_O \left( \langle O \rangle (\boldsymbol{\theta}) - \langle O \rangle_\mathrm{exp} \right)^2,
  \label{eq:loss-obs}
\end{equation}
where $p_O$ is the weight of the new term and $\langle O \rangle_\mathrm{exp}$ the experimental value of the observable.
Eq.~\eqref{eq:loss-obs} can easily be generalized to the case of multiple observables by summing over their respective mean square errors.
Generating a model using this loss function poses a significant problem, namely, computing $\langle O \rangle (\boldsymbol{\theta})$ on the fly during the training.
This quantity is impractical to calculate via Eq.~\eqref{eq:ensemble-obs}, since this would require stopping the training process to launch MD simulations.
Therefore, we require a methodology to compute $\langle O \rangle (\boldsymbol{\theta})$ efficiently and on the fly during training.

To solve this conundrum, we resort to reweighting, which is an exact and well-known method to compute ensemble averages of an observable under the action of a different PES~\cite{torrie-1977,piaggi-2019}. Using this approach, we obtain that the ensemble average of the observable for a new model is given by,
\begin{equation}
  \langle O \rangle (\boldsymbol{\theta}) = \frac{\sum_{i=1}^N O_i\, e^{-\beta \left( E(\mathbf{R}^i;\boldsymbol{\theta}) - E(\mathbf{R}^i;\boldsymbol{\theta}_0) \right)}}{\sum_{i=1}^N e^{-\beta \left( E(\mathbf{R}^i;\boldsymbol{\theta}) - E(\mathbf{R}^i;\boldsymbol{\theta}_0) \right)}},
  \label{eq:reweighting}
\end{equation}
where $\boldsymbol{\theta}_0$ denotes the parameters of the reference model (used to generate the MD trajectory), and $\boldsymbol{\theta}$ are the parameters of the new model whose ensemble average we aim to estimate.

Before moving forward to discuss our algorithm, we make a few remarks. First, similar strategies have been suggested in the past for fitting simpler models for the PES to empirical data~\cite{norgaard2008experimental,frohlking2020toward}, yet here we take advantage of the machinery of artificial intelligence to treat both \textit{ab initio} and empirical data on the same footing, and to learn them simultaneously.
Also, the gradient $\nabla_\theta\mathcal{L}(\boldsymbol{\theta})$, needed for gradient-descent-like optimization methods, is well defined and is trivial to implement in modern machine learning frameworks with automatic differentiation.
Finally, we would like to point out the fundamental difference between the force loss term, which aims to learn an atomic property within a single microstate, and the observable loss term, which instead learns highly collective properties that are only defined for the macrostate (multiple microstates). Therefore, the effect of learning the experimental value of an observable is distributed among all atoms and over many different configurations, ensuring a smooth and subtle modification of the \textit{ab initio} forces.

\begin{figure}[t]
  \centering
  \includegraphics[width=\columnwidth]{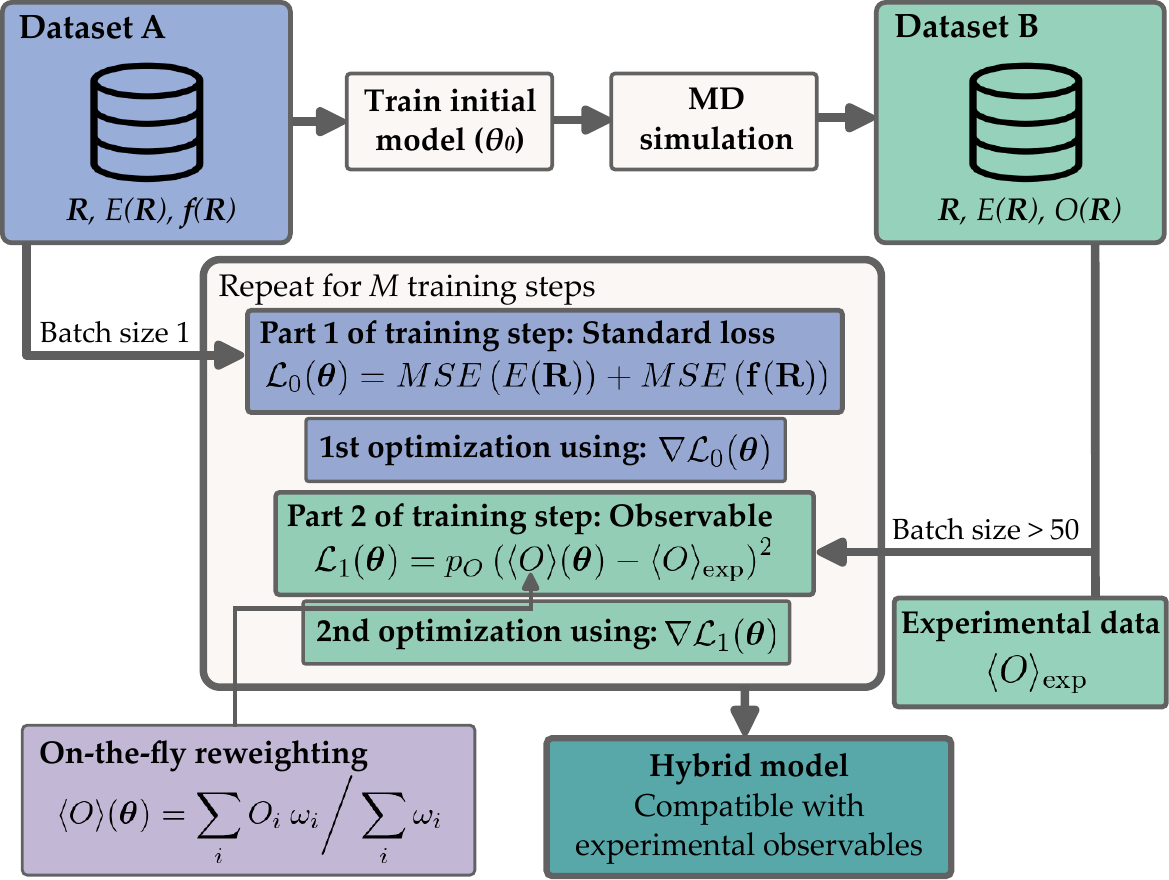}
  \caption{Schematic architecture of our method. Dataset A contains atomic coordinates, and \textit{ab initio} energies and forces. Dataset B is extracted from an MD simulation driven by a model trained with Dataset A, and contains atomic coordinates, energies, and instantaneous values of observables. Each iteration of the training process is divided into two parts: in part 1, energies and forces are learned from dataset A using their mean square error (MSE) as loss function, whereas in part 2, the ensemble average of the observable is computed using on-the-fly reweighting with weights $\omega_i=e^{-\beta \left( E(\mathbf{R}^i;\boldsymbol{\theta}) - E(\mathbf{R}^i;\boldsymbol{\theta}_0) \right)}$ as shown in Eq.~\eqref{eq:reweighting}, and its MSE with respect to the experimental data is optimized. After $M$ trainings steps, we obtain a hybrid model compatible with both the \textit{ab initio} and experimental data.}
  \label{fig:fig1}
\end{figure}

Now that the theoretical underpinnings of our method have been fully described, we turn our attention to the design of our algorithm, which is shown schematically in Fig.~\ref{fig:fig1}.
Our method uses two training datasets, labeled A and B. Dataset A contains atomic configurations of the system (i.e., atomic coordinates and cell matrices) $\mathbf{R}$ along with their corresponding energies $E_\mathrm{ref}^i$ and forces $\mathbf{f}_\mathrm{ref}^i$, computed using electronic structure calculations. This training set contains the \textit{ab initio} part of the input data and typically contains between $10^3$ and $10^5$ atomic configurations. With dataset A we train a model for the PES based on the standard loss function in Eq.~\eqref{eq:loss-ef}, with final training parameters $\boldsymbol{\theta}_0$. Using this model, we run an MD simulation in the isothermal--isobaric ensemble which will be used to create dataset B. This new dataset contains between $10^3$ and $10^4$ atomic configurations sampled from the MD simulation with their corresponding energies $E(\mathbf{R};\boldsymbol{\theta}_0)$, and the instantaneous value of the observable $O(\mathbf{R})$ we aim to fit.

The training is an iterative process and each iteration is divided into two parts. In the first part, the model parameters $\boldsymbol{\theta}$ are optimized using the gradient of the standard loss function for energies and forces in Eq.~\eqref{eq:loss-ef}. This stage makes use of dataset A, as it focuses on learning the PES only from \textit{ab initio} data. On the other hand, the second part of the iteration uses dataset B and focuses on optimizing the model parameters $\boldsymbol{\theta}$ to reproduce the experimental value of the observable. For this purpose, we use the loss function in Eq.~\eqref{eq:loss-obs}. The ensemble average $\langle O \rangle (\boldsymbol{\theta})$ is computed on the fly in each step through the reweighting formula in Eq.~\eqref{eq:reweighting}. An important difference between these two steps is that while the first part uses a mini-batch of size 1, i.e., only one configuration is used in each step, the second part needs a larger mini-batch for a proper calculation of the ensemble average.

Before discussing our results, we pause briefly to describe the computational details of our calculations.
The methodology used for constructing the models for the PES is based on the Deep Potential (DeePMD) approach developed by Zhang \textit{et al.}~\cite{zhang-2018, zhang-2018-smooth}, and the implementation was built on top of the DeePMD-JAX~\cite{gao-2024} package. We chose to work with this software as it was specifically designed for fast development and testing. In our calculations, the energy was represented by a neural network with three layers and 128 neurons per layer, and the embedding matrix (descriptors) was represented by a three-layer neural network with two layers of size 32 and one of size 64. Interactions were truncated at a cutoff $r_c = 6 \: \text{\AA}$. Other details of the model can be found in Ref.~\cite{gao-2024}.
The parameters in the training procedure were as follows. The models were trained for $M=10^6$ iterations using the Adam optimizer with an exponential decay every $5\times 10^3$ steps and a learning rate $l_r(i)$ going from $2\times10^{-3}$ to $10^{-6}$ as a function of the iteration step $i$. The prefactors in the loss function also varied according to $p(i) = p^f + (p^s-p^f) \: l_r(i)/l_r(0)$, where $p^s$ and $p^f$ are the starting and final prefactors, respectively. The starting values were $p_E^s=0.02$, $p_f^s=1000$ and $p_O^s=0.2$; and the end values $p_E^f=1$, $p_f^f=1$ and $p_O^f=100$.
The mini-batch sizes were 1 and 100 for parts 1 and 2 of the iteration, respectively. In the fitting of the radial distribution functions we increased the batch size for part 2 to 500.
MD simulations were carried out using the JAX-MD~\cite{schoenholz2020jax} code. The integration of the equations of motion used the Suzuki--Trotter integrator~\cite{suzuki-1991}, with a time step of 0.5~fs. We used the standard atomic weights for the masses of oxygen and hydrogen. All simulations were ran at isothermal-isobaric conditions, with the pressure set to 1 bar. We controlled the temperature and pressure using the Nosé--Hoover thermostat~\cite{hoover-1985} and barostat~\cite{martyna-1994} with a relaxation time of 0.1~ps and 1~ps, respectively.

\begin{figure}[t]
  \centering
  \includegraphics[width=\columnwidth]{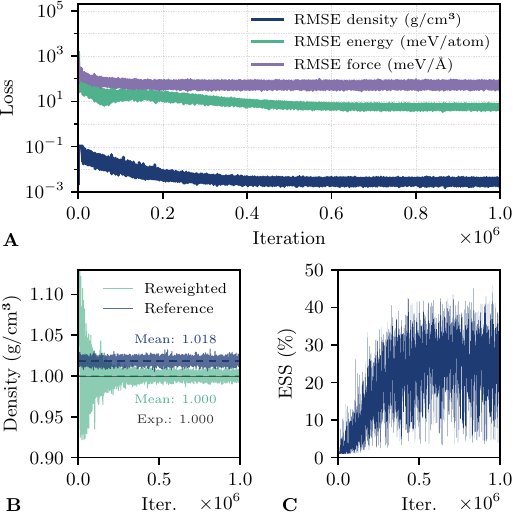}
  \caption{Training process of a hybrid model for liquid water consistent with the experimental density. A) Evolution of each term in the loss function during the training process, represented using the root mean square error (RMSE). B) Non-reweighted mean (reference) and reweighted mean of the density computed over a mini-batch of 100 configurations during training. C) Effective sampling size (ESS), which quantifies the efficacy of the reweighting process (see text for details).}
  \label{fig:fig2}
\end{figure}

\begin{figure}[t]
  \centering
  \includegraphics[width=\columnwidth]{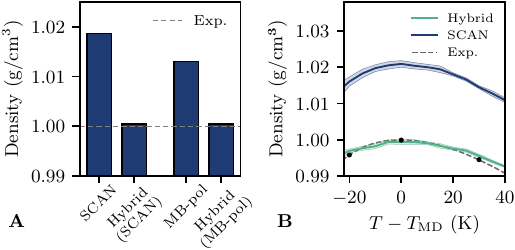}
  \caption{Density of water from MD simulations driven by purely \textit{ab initio} and hybrid models. A) Density at the temperature of maximum density $T_\mathrm{MD}$ for models based on SCAN ($T_\mathrm{MD}=320$ K) and MB-pol ($T_\mathrm{MD}=260$ K). The hybrid models are trained using the experimental density at $T_\mathrm{MD}=297$~K, which is shown as a horizontal dashed line~\cite{nistwebbook}. B) Density isobar at 1 bar for purely \textit{ab initio} and hybrid models based on SCAN. The hybrid model is trained on the three experimental densities shown with black dots. The experimental data is taken from Refs. \citenum{nistwebbook} and \citenum{hare1987density}. The shaded areas around each curve represent the error estimated using block averages.}
  \label{fig:fig3}
\end{figure}

We first describe the application of our algorithm to develop a model for liquid water consistent with the experimental density $0.99995\ \mathrm{g/cm^3}$ at 277 K and 1 bar~\cite{nistwebbook}, which corresponds to the density maximum.
We use the \textit{ab initio} dataset described in Ref.~\cite{zhang-2021}, and consisting of energies and forces computed using the Strongly Constrained and Appropriately Normed (SCAN) DFT functional~\cite{SCAN-2015}.
This dataset covers liquid water and multiple ice polymorphs at thermodynamic conditions spanning a large swath of the temperature-pressure phase diagram.
We trained a model purely based on this \textit{ab initio} dataset and found that it has a temperature of maximum density $T_\mathrm{MD}$ of 320 K and the corresponding density is $1.018\ \mathrm{g/cm^3}$, consistent with previous studies using the same dataset~\cite{piaggi2021phase}, yet somewhat lower than estimates using direct \textit{ab initio} MD~\cite{chen2017ab}.
We then constructed dataset B, which is used for fitting the experimental density, by carrying out a simulation at the $T_\mathrm{MD}$ for this model.
Equipped with these two datasets, we now turn to discuss the results of the training of our hybrid model.
Fig.~\ref{fig:fig2}A shows the evolution of the different terms of the loss function during the training process.
The root mean square errors (RMSEs) for energy and forces decrease progressively, and are smaller than $8\ \mathrm{meV/atom}$ and $80\ \mathrm{meV/\text{\AA}}$, respectively, by the end of the training process.
These results are consistent with previously reported training processes using DeePMD and a similar dataset~\cite{piaggi-2024}.
A substantial decrease is also observed in the observable loss term, reaching values smaller than $0.004\ \mathrm{g/cm^{3}}$.
Overall, the results show the expected evolution of a successful learning process.
In Fig.~\ref{fig:fig2}B, we show the non-reweighted mean value of the density and the reweighted mean from Eq.~\eqref{eq:reweighting} for each iteration step.
The average non-reweighted density is $1.018\ \mathrm{g/cm^3}$, as reported above for the purely \textit{ab initio} model, while the reweighted density is $1.000\ \mathrm{g/cm^3}$, with a negligible error with respect to the experimental counterpart.
A useful way to measure the effectiveness of the reweighting process is by using the Effective Sample Size (ESS)~\cite{invernizzi2020unified}, defined as $(\sum_{i=1}^N \omega_i)^2 / \sum_{i=1}^N \omega_i^2$, where $\omega_i = e^{-\beta \left( E^i(\mathbf{R};\boldsymbol{\theta}) - E^i(\mathbf{R};\boldsymbol{\theta}_0) \right)}$ is the weight of the $i$-th configuration of the mini-batch, and $N$ is the size of the mini-batch. The ESS is 1 when there is a single non-zero weight, and $N$ if all the weights are equal. As shown in Fig.~\ref{fig:fig2}C, the ESS starts at around zero in the initial stages of the training due to the inability of a random model to yield an effective reweighted estimate. However, after $\sim 4 \times 10^5$ iterations a sensible solution is found and the ESS increases to around 10-40 \%, indicating a successful estimate of $\langle O \rangle (\boldsymbol{\theta})$.

With the hybrid model successfully trained, we computed the density using MD driven by this model and the result is shown in Fig.~\ref{fig:fig3}A.
The density at $T_\mathrm{MD}$ was reduced from $1.018\ \mathrm{g/cm^3}$ in the purely \textit{ab initio} SCAN model to $1.000\ \mathrm{g/cm^3}$ in the hybrid model, showing the ability of the technique to create models with a precise density, which agrees with experimental data.
In order to test the robustness of the methodology for different datasets, we repeated the training described above using the dataset reported in Ref.~\cite{bore2023realistic}, derived from the MB-pol model of water which is based on highly-accurate coupled-cluster CCSD(T) calculations~\cite{babin-2013}.
In Fig.~\ref{fig:fig3}A, we show the densities of a model purely based on the MB-pol dataset and of a hybrid model that targets the density at $T_\mathrm{MD}$, which is 260 K for this model.
Also in this case, our methodology reduces the density from $1.013\ \mathrm{g/cm^3}$, which is somewhat better than SCAN, to $1.000\ \mathrm{g/cm^3}$ in perfect agreement with the experiment.
So far, we have focused on studying the behavior of the density at a single temperature.
In Fig.~\ref{fig:fig3}B, we compare the results of the density isobar at 1 bar for the model purely based on SCAN with a hybrid model that targets the experimental density at three different temperatures.
The hybrid model closely follows the experimental data across the studied temperature range.

\begin{figure}[t]
  \centering
  \includegraphics[width=\columnwidth]{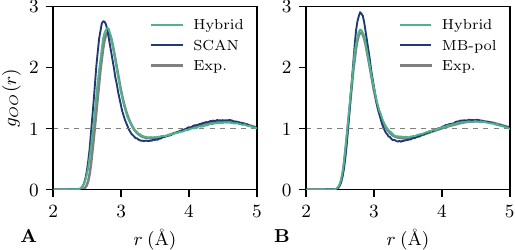}
  \caption{Radial distribution functions $g_{OO}(r)$ at temperature $T=T_\mathrm{MD}+18$~K from MD simulations driven by purely \textit{ab initio} and hybrid models based on A) SCAN and B) MB-pol. The hybrid models were trained on the experimental $g_{OO}(r)$ at $T=295$~K, which is also shown in the plots and was taken from Ref.~\cite{skinner2013benchmark}.}
  \label{fig:fig4}
\end{figure}

We now test our technique for the task of creating a model with an O-O radial distribution function $g_{OO}(r)$ consistent with experiment. The $g_{OO}(r)$ is a stringent test of the ability of \textit{ab initio} methods to describe the structure of the liquid, which depends on the subtle interplay between different types of interactions, including van der Waals and hydrogen bonding.
In Fig.~\ref{fig:fig4}A, we show the experimental $g_{OO}(r)$ at 295~K, which is 18~K above $T_\mathrm{MD}$, together with the $g_{OO}(r)$ for the SCAN model at the corresponding temperature compared to its $T_\mathrm{MD}$, namely, 338~K.
The model based on SCAN has the first O-O peak shifted to a shorter distance compared to the experimental counterpart and is also somewhat overstructured.
We then trained a hybrid model based on the SCAN dataset and using the experimental $g_{OO}(r)$ for $2.25$~\AA~$<r<5$~\AA~with a spacing of 0.01~\AA.
Note that each bin is considered a different observable and each of them contributes a term to Eq.~\eqref{eq:loss-obs}.
The $g_{OO}(r)$ for the hybrid model thus trained is shown in Fig.~\ref{fig:fig4}A and we found excellent agreement with the experimental $g_{OO}(r)$, except for some very small differences at the first peak.
In Fig.~\ref{fig:fig4}B, we repeated this comparison using the dataset based on MB-pol at the corresponding temperature, namely 278~K, with similar results.
The model based solely on MB-pol describes better than SCAN the position of the first peak of the $g_{OO}(r)$, yet its height is overestimated.
It must be borne in mind that nuclear quantum effects decrease the height of this peak bringing it into better agreement with experiment~\cite{medders2014development}.
The hybrid model based on MB-pol reduces the discrepancy and its performance is similar to the hybrid model based on SCAN.










In summary, we presented a methodology that provides a general way of obtaining hybrid machine learning models trained simultaneously on \textit{ab initio} and empirical data.
We have shown that this approach can produce models that retain the main characteristics of the \textit{ab initio} PES, while improving the predictions of one or more observables, with respect to experiment.
We also stress that, besides correcting inaccuracies in electronic structure calculations, our method can also be used to incorporate nuclear quantum effects on structural or thermodynamic properties in an effective fashion, circumventing the need to carry out path integral MD, in some cases.
Finally, the results reported here will stimulate future work to disseminate and extend this new methodology.

\section*{Data Availability Statement}

All data needed to reproduce our results, including software, datasets, training scripts, molecular dynamics inputs, and analysis scripts, are openly available on Zenodo (DOI \url{https://doi.org/10.5281/zenodo.17610189})~\cite{dataset}.

\section*{Acknowledgements}
P.PC. acknowledges the generous support provided by the Ikasiker scholarship from the Basque Government and nanoGUNE's master scholarship.
P.M.P. acknowledges funding from the Marie Skłodowska-Curie Cofund Programme of European Commission project H2020-MSCA-COFUND-2020-101034228-WOLFRAM2.
We are grateful for the computational resources provided on the Hyperion cluster at the Donostia International Physics Center (DIPC) and on MareNostrum 5 ACC at the Barcelona Supercomputing Center, granted through the Spanish Supercomputing Network (RES) allocations FI-2024-2-0026 and FI-2024-3-0028.

\bibliography{bibliography}

\end{document}